\documentclass[twocolumn,letterpaper,preprintnumbers,amsmath,amssymb,nofootinbib]{revtex4-1}
\pdfoutput=1
\usepackage{amssymb}
\usepackage{graphicx}
\usepackage{amsmath}
\usepackage{multirow}
\usepackage{verbatim}
\usepackage{flushend}
\usepackage[usenames,dvipsnames]{color}
\usepackage[bookmarks=true, bookmarksnumbered=true, colorlinks=true, urlcolor=darkblue, citecolor=darkgreen, linkcolor=darkred, breaklinks=true]{hyperref}

\definecolor{darkblue}{rgb}{0,0,0.5}
\definecolor{darkgreen}{rgb}{0.1,0,0.3}
\definecolor{darkred}{rgb}{0.6,0,0}

\newcommand{\nc}{\newcommand}
\nc{\ba}{\begin{eqnarray}}
\nc{\ea}{\end{eqnarray}}

\newcommand\s{\sigma}

\nc{\ga}{\gamma}
\nc{\om}{\omega}

\nc{\x}{{\bf x }}
\nc{\kk}{{\bf k }}
\nc{\f}{{\bf f }}
\nc{\e}{{\bf e }}
\nc{\T}{ \theta (s_i (t)- \s) }
\nc{\TT}{ \theta (s_i (t_{ r \, i } )- \s) }
\nc{\br}{   (s_i (t)- \s)  }
\nc{\gta}{\gamma \rightarrow a}
\nc{\D}{\Delta}
\nc{\Dag}{\Delta_{a \gamma}}
\nc{\Dosc}{\Delta_{\rm osc}}
\nc{\Dpl}{\Delta_{\rm pl}}
\nc{\Da}{\Delta_a}
\nc{\gag}{g_{a \gamma}}
\nc{\wpl}{\omega_{\rm pl}}
\nc{\hr}{$h_{\rm res}$}
\nc{\ud}{\mathrm{d}}
\nc{\igev}{GeV$^{-1}$}
\nc{\eg}{\textit{e.g.}}
\nc{\Hk}{H_{\kappa}}

\begin{document}

%%%%%%%%%%%%%%%%%%%%%%%%%%%%%%%%%%%%%%%%%%%%%%%%%%%%%%%%%%%%%%%%%%%%%%%
\title{Light bosons and photospheric solutions to the solar abundance problem}

\author{Aaron C. Vincent}
\email{vincent@ific.uv.es}
\affiliation{IFIC, Universitat de Val\`encia --- CSIC, 46071, Valencia, Spain}
\author{Pat Scott} 
\email{patscott@physics.mcgill.ca}
\affiliation{Department of Physics, McGill University 3600 Rue University, Montr\'eal, Qu\'ebec, Canada H3A 2T8}
\author{Regner Trampedach}
\email{trampeda@lcd.colorado.edu}
\affiliation{JILA, University of Colorado and National Institute of Standards and Technology, 440 UCB, Boulder, CO 80309, USA}

\begin{abstract}
It is well known that current spectroscopic determinations of the chemical composition of the Sun are starkly at odds with the metallicity implied by helioseismology.  We investigate whether the discrepancy may be due to conversion of photons to a new light boson in the solar photosphere.  We examine the impact of particles with axion-like interactions with the photon on the inferred photospheric abundances, showing that resonant axion-photon conversion is not possible in the region of the solar atmosphere in which line-formation occurs. Although non-resonant conversion in the line-forming regions can in principle impact derived abundances, constraints from axion-photon conversion experiments rule out the couplings necessary for these effects to be detectable. We show that this extends to hidden photons and chameleons (which would exhibit similar phenomenological behaviour), ruling out known theories of new light bosons as photospheric solutions to the solar abundance problem.  
\end{abstract}

\maketitle

\section{Introduction}
Agreement between predictions of solar interior models and helioseismological measurements has deteriorated in recent years, due to a downwards revision in standard solar abundances.  Recent photospheric analyses \cite{APForbidO,CtoO,AspIV,AspVI,AGS05,ScottVII,Melendez08,Scott09Ni,AGSS} have indicated that the solar metallicity is about $20\%$ lower than previously thought \cite{GS98}. In particular, the current reference photospheric abundances \cite{AGSS} of the important elements C, N and O are smaller than the previous reference values \cite{GS98} by 0.09, 0.09 and 0.14\,dex, respectively.  Standard solar models computed with the revised surface metal abundances show large discrepancies with the depth of the convection zone, the surface helium abundance and the sound speed profile inferred from helioseismology \cite{Bahcall:2004yr, Basu:2004zg, Bahcall06, Yang07, Basu08, Serenelli:2009yc}.  This has been variously referred to as `the solar model(ling) problem' \cite{Asplund08,Drake05,Morel08}, `the solar oxygen crisis' \cite{Ayres06,Ayres08,Socas07}, or `the solar abundance problem' \cite{Guzik10,Serenelli11}.

The most common approaches to the solar abundance problem to date have been to recheck the photospheric abundances \cite{Drake05, Ayres06, Ayres08, Morel08, Caffau08, Centeno08, Caffau10}, or to modify solar models in an attempt to accommodate both the measured abundances and data from helioseismology. Efforts in the latter category have included enhanced diffusion \cite{Guzik05}, late-stage accretion of low-metallicity gas \cite{Guzik05,Castro07,Guzik10,Serenelli11}, modifications of opacities \cite{Bahcall05, Badnell05, Christensen09}, internal gravity waves \cite{Arnett05,Charbonnel05} and enhanced energy transport in the solar core due to gravitational capture and scattering of dark matter \cite{Frandsen10,Taoso10,Cumberbatch10}.  Despite nearly a decade of effort (and a few errant claims to the contrary), none of these approaches has proven successful in solving the problem.

Here we propose an alternative approach: examining effects of new physics on the measurement of solar photospheric abundances. Specifically, we investigate the impact on derived elemental abundances of mixing between photons and new light bosons, such as axions, in the line-forming regions of the photosphere.\footnote{A more complicated but related scenario was considered in \cite{Zioutas:2007xk}, where high-energy axions produced in the solar core would reconvert to photons in the photosphere and exert radiation pressure preferentially on heavier elements, driving them from the photosphere.}

The 5770\,K thermal spectrum of the Sun is formed around optical depth $\tau = 1$, which defines the location of the solar `surface'.  For a height $z$ above the surface,
\begin{equation}
\tau_\lambda(z) \equiv \int_z^\infty \alpha_\lambda(z') \ud z'.
\end{equation}
Here $\alpha_\lambda(z) \equiv \kappa_\lambda^{\rm c}(z) \rho(z)$ is the linear attenuation coefficient at wavelength $\lambda$, where $\kappa_\lambda^{\rm c}$ is the continuum opacity and $\rho$ is the local gas density. Solar abundances are inferred from absorption lines. These are formed as atomic or molecular bound-bound transitions increase the opacity in small wavelength ranges around the transitions. The higher opacity in these spectral lines means that the light at those wavelengths escapes from layers above the continuum-forming layers (the photosphere). As the gas temperature decreases with height, the line-centre photons escape from cooler layers than the continuum photons, giving rise to absorption lines. 

The equivalent width of a line (the integrated area between the line and continuum) is a relative measure of line strength, and depends not only on the abundance of a given element, but also on the attenuation of the total continuum as it passes through the line-forming region. For this reason, both line and continuum fluxes must be carefully calculated with radiative transfer simulations before elemental abundances can be inferred.  If additional plasma effects in the photosphere enhance continuum photon attenuation via oscillation into light bosons such as axions, then the equivalent width of absorption lines used to calculate elemental abundances would be suppressed. By combining the axion conversion effect with earlier radiative transfer calculations, we will show how this could occur in principle, and assess whether this effect could arise from existing particle theories in such a way as to solve the solar abundance problem. 

Historically, the axion field arose as a candidate solution to the strong CP problem of QCD. Although the term 
\begin{equation}
  \tilde \Theta_3 G_{\mu\nu}\tilde G^{\mu \nu},
\label{CPviol}
\end{equation} 
is not forbidden in any way in the QCD Lagrangian, it seems to be absent from the standard model (SM) of particle physics.  Here $G^{\mu\nu}$ is the gluon field strength, and its dual is $\tilde G^{\mu\nu} = \epsilon^{\alpha \beta \mu \nu} G_{\alpha \beta}$. The coupling constant $\tilde \Theta_3$ is experimentally consistent with zero, so some sort of dynamical method is required in order to force it to be small without fine-tuning. The Peccei-Quinn (PQ; \cite{PQ}) mechanism introduces a broken U(1) symmetry that naturally yields this cancellation, with the added effect of producing a new light, propagating, pseudoscalar degree of freedom: the axion. 

The proposed existence of the axion field has given rise to a rich variety of phenomenological predictions. These include axion production in the dense cores of stars (\cite{Raffelt19901} and references therein), ``light shining through wall'' effects \cite{vanBibber1987}, and many theories of axionic dark matter (\cite{Duffy:2009ig} and references therein). Many of these effects are due to the effective coupling $\mathcal{L}_{a\gamma \gamma}$ that must arise between the PQ axion field $a$ and the photon gauge field strength $F^{\mu \nu}$:
 \begin{equation}
\mathcal{L}_{a\gamma \gamma} =\frac{\gag}{4}a F_{\mu \nu} \tilde F^{\mu \nu} = -\gag a \boldsymbol{E} \cdot \boldsymbol{B}\,
\label{axionFF}
\end{equation} 
with coupling constant $\gag$. It is this axion-photon-photon vertex that allows axions to convert into photons in the presence of electric ($\boldsymbol{E}$) or magnetic ($\boldsymbol{B}$) fields, and vice versa. The mass and coupling of a PQ axion must be very small (with $m_a \propto 1/g_{a\gamma}$), and currently remain (mostly) far below the threshold of experimental searches. A rough lower limit of 10$^{-6}$\,eV can but put on the axion mass, in order to avoid overclosing the universe, and an upper limit of $m_a \lesssim 10^{-3}$\,eV ensures that stellar evolution is not significantly modified \cite{Asztalos:2006kz}.

The PQ mechanism in QCD is not the only possible origin of such a particle, however. Axion-like particles (ALPs) are a ubiquitous element in string theory compactifications \cite{Svrcek:2006yi,Dasgupta:2008hb}, in which the moduli describing the compact extra dimensions can appear to observers as internal $U(1)$ gauge degrees of freedom.  More generally, $U(1)$ symmetries appear in many high-energy extensions of the SM, giving rise to exotic particles with interactions similar to the axion.  Like axions, ALPs interact with photons via a term of the form Eq.~\ref{axionFF}.  Unlike axions, they do not necessarily solve the strong CP problem, nor require such small couplings or $m_a \propto 1/g_{a\gamma}$.

We begin by showing in Section \ref{sec:theory} how photon-ALP oscillation can affect the linear attenuation coefficient -- and hence the equivalent width of an absorption line -- in the solar atmosphere at optical depths $\tau < 1$. We demonstrate in Section \ref{sec:abundances} that with a large enough coupling $\gag$, a low-mass particle with a term like Eq.~\ref{axionFF} could affect the solar atmosphere enough to modify absorption lines in an observable way. As we discuss in Section \ref{sec:discussion} however, the coupling strength required to produce such an effect is much larger than the coupling allowed by current experimental bounds on standard ALPs.  We go on to show that this rules out not only ALPs as photospheric solutions to the solar abundance problem, but also hidden photons and chameleons.  We conclude in Section \ref{sec:conclusion} that unless some other light boson with an axion-like or similar kinetic-mixing type coupling to the photon can be found that somehow evades the existing experimental constraints, the solar abundance problem cannot be solved by light bosons in the photosphere.

\section{Theory}
\label{sec:theory}
\subsection{Changing inferred abundances}
\label{sec:theoryA}

We first consider the standard case, without ALPs.  On the linear portion of the curve of growth, the equivalent width of an absorption line $W_\lambda$ is approximately proportional to the local ratio of the line-centre ($\kappa_\lambda^\mathrm{l}$) to continuum ($\kappa_\lambda^\mathrm{c}$) opacities at the height of line formation \cite{Rutten}:
\begin{equation}
\label{EWnormal}
W_\lambda \varpropto \frac{\kappa_\lambda^\mathrm{l}}{\kappa_\lambda^\mathrm{c}}\equiv \eta_\lambda.
\end{equation}
The abundance-dependence of $W_\lambda$ is encoded in $\eta_\lambda$ by way of $\kappa_\lambda^\mathrm{l}$, which is itself proportional to the number of absorbers along the line of sight.  The number of absorbers is given by the number of atoms with electrons in the lower level of the transition, which is some known (temperature- and density-dependent) fraction of the total abundance of the element in question, $\epsilon$.  

In the presence of ALPs, the local continuum opacity receives an additional effective contribution $\kappa_\lambda^\mathrm{a}$ from photon-ALP conversion.  In this case
\begin{equation}
\label{EWaxion}
W_\lambda \varpropto \frac{\kappa_\lambda^\mathrm{l}}{\kappa_\lambda^\mathrm{c}+\kappa_\lambda^\mathrm{a}}= \frac{\eta_\lambda}{1 + \eta_\lambda^\mathrm{a}},
\end{equation}
where we have defined the `axionic extinction ratio' $\eta_\lambda^\mathrm{a}\equiv\kappa_\lambda^\mathrm{a}/\kappa_\lambda^\mathrm{c}$ in analogy with the line extinction ratio $\eta_\lambda$.  The abundance-dependence of Eq.~\ref{EWaxion} is identical to that of Eq.~\ref{EWnormal},\textit{ i.e. }linear in $\epsilon$, with the same constant of proportionality.  However, given that $W_\lambda$ is known from observation, analysis using either Eq.~\ref{EWnormal} or Eq.~\ref{EWaxion} is constrained to return the same answer for $W_\lambda$.  Denoting the abundance derived by ignoring ALPs (Eq.~\ref{EWnormal}) as $\epsilon_\mathrm{std}$, and that including the effects of ALPs (Eq.~\ref{EWaxion}) as $\epsilon_\mathrm{corr}$, we have 
\begin{equation}
    \frac{W_\lambda\ ({\rm Eq.~\hbox{\ref{EWnormal}}})}
         {W_\lambda\ ({\rm Eq.~\hbox{\ref{EWaxion}}})}
  = \frac{\epsilon_\mathrm{std}}
         {\epsilon_\mathrm{corr}/(1 + \eta_\lambda^\mathrm{a})}=1.
\end{equation}
The abundance correction to be applied to the standard result in order to obtain the ALP-corrected abundance is thus 
\begin{equation}
\Delta\log\epsilon \equiv \log\epsilon_\mathrm{corr} - \log\epsilon_\mathrm{std} = \log(1+\eta_\lambda^\mathrm{a}).
\label{Dloge}
\end{equation}
With Eq.~\ref{Dloge} in hand, the remaining task is then to calculate $\eta_\lambda^\mathrm{a}$ in the line-forming regions of the photosphere.

\subsection{Photon-ALP conversion in the Solar atmosphere}
\label{sec:theoryB}

Regardless of the precise ALP model, the presence of a coupling term with the photon (Eq.~\ref{axionFF}) implies that under the right conditions, a thermal photon in the photosphere may oscillate into an ALP and vice-versa. This occurs in the presence of an external photon field: either a bulk magnetic field, or the oscillating fields of the stellar plasma. Only the former effect is relevant in the region of interest for line formation, as plasma field-induced transition rates (Eq.~\ref{gammaTemp}) turn out to be extremely small in the photosphere. In order to arrive at the quantity $\eta^a_\lambda$ we first compute the linear attenuation coefficient $\alpha^a_\lambda$, related to the ALP opacity by:
\begin{equation}
 \alpha^a_\lambda \equiv \kappa_\lambda^a \rho.
\end{equation}

\subsubsection{Conversion in plasma fields}
The transition from photon $\gamma$ to ALP $a$ induced by the oscillating electric fields in a plasma occurs mainly from resonant Primakoff conversion. Neglecting recoils, the transition rate is given by \cite{Raffelt:1988, Raffelt:2006cw} :
\begin{equation}
 \alpha_{\lambda,\mathrm{pl}}^a = \frac{\gag^2 T \kappa_s^2}{32\pi}\left[\left(1+\frac{\kappa_s^2}{4\omega^2} \right) \ln\left(1+\frac{4\omega^2}{\kappa_s^2}\right) \right]
\label{gammaTemp}
\end{equation} 
where $\kappa_s$ is the screening scale in the Debye-H\"uckel approximation:
\begin{equation}
 \kappa_s^2 = \frac{4\pi\alpha}{T}\left(n_{\rm e} + \sum_{\rm nuclei}Z_{\rm i}^2n_{\rm i}\right),
\end{equation} 
$Z_{\rm i}$ is the charge of the $i^{\rm th}$ ion and the number densities $n_{\rm e}$, $n_{\rm i}$ of electrons and ions, respectively, are found by solving the equation of state. This effect is highly suppressed at the relatively low charge densities of the solar atmosphere, and contributes a negligible amount at optical depths of interest to us. 

\subsubsection{Conversion in a bulk magnetic field}

Conversion in a bulk magnetic field can be much more significant than in the plasma field. The probability that a photon polarised in the direction of the $\boldsymbol{B}$ field will have converted to an ALP after travelling a distance $\delta z \equiv z - z_0$ from height $z_0$ to height $z$ is \cite{Raffelt:1987im, Mirizzi:2006zy}:
\begin{equation}
 P_0(\delta z) = 4\left(\frac{\Dag}{\Dosc}\right)^2 \sin^2(\Dosc \delta z/2)
\label{P0}
\end{equation} 
where:
\begin{align}
\Dosc^2 &= (\D_{||}-\Da)^2 + 4\Dag^2, \\
 \D_{||} & = \D_{||}^{\rm{vac} }+ \D_{||}^{\rm{gas}} + \Dpl \nonumber \\
 \Da &= -\frac{m_a^2}{2\omega}	\nonumber\\
 \Dag &= \gag |\boldsymbol{B}_T|/2,	\nonumber\\
\Dpl &= -\frac{\wpl^2}{2\omega}.\nonumber
\end{align} 
$\boldsymbol{B}_T$ is the component of the magnetic field transverse to the direction in which the photon propagates, $\omega=2\pi c/\lambda$ is the frequency of the incoming photon, and the plasma frequency is $\wpl = 4\pi\alpha n_{\rm e}/m_{\rm e}$, where $\alpha$ is the fine-structure or electromagnetic coupling constant. Due to the $\boldsymbol E\cdot \boldsymbol B$ form of the interaction (\ref{axionFF}), only the photons with polarisation parallel to the magnetic field are susceptible to conversion; this is the meaning of the $||$ subscript. $\D_{||}^{\rm{vac}} = \omega (n_{||}^{\mathrm{vac}}-1) $, where $n_{||}^{\mathrm{vac}} \sim 1 + (\alpha B/4\pi B_{\mathrm{c}})$ is the ``vacuum'' contribution to the refractive index induced by the magnetic field. The critical field $B_c = m_e^2/e \sim 10^{12}\,\mathrm{eV}^2 \sim 4 \times 10^9$\,T is much larger than the fields $B \sim 0.1$\,T that we consider in this work; $\D_{||}^{\rm{vac}}$ is therefore negligible for all cases of interest here.  Similarly, $\D_{||}^{\rm{gas}} = \omega(n_{||}^{\rm{gas}} -1)$, where $n_{||}^{\rm{gas}}$ is the contribution to the refractive index by neutral hydrogen gas. 
A relativistic calculation of the static polarizability of hydrogen \cite{bartlett:HIpolariz} gives $\alpha_0(\mathrm{H}\textsc{i}) = 6.6679437 \times 10^{-25} \, \mathrm{cm}^3$. The index of refraction is then:
\begin{equation}
n_{||}^{\rm{gas}} = \sqrt{1 + 4\pi n(\mathrm{H}\textsc{i})\alpha_0(\mathrm{H}\textsc{i})}
\end{equation}
where $n(\mathrm{H}\textsc{i})$ is the number density of neutral hydrogen atoms. This is valid in the low densities of the solar atmosphere and as long as we are sufficiently far from the Lyman\,$\alpha$ line. 

\begin{figure}[tbp]
\hspace{-0.4cm}
\includegraphics[width=.5\textwidth]{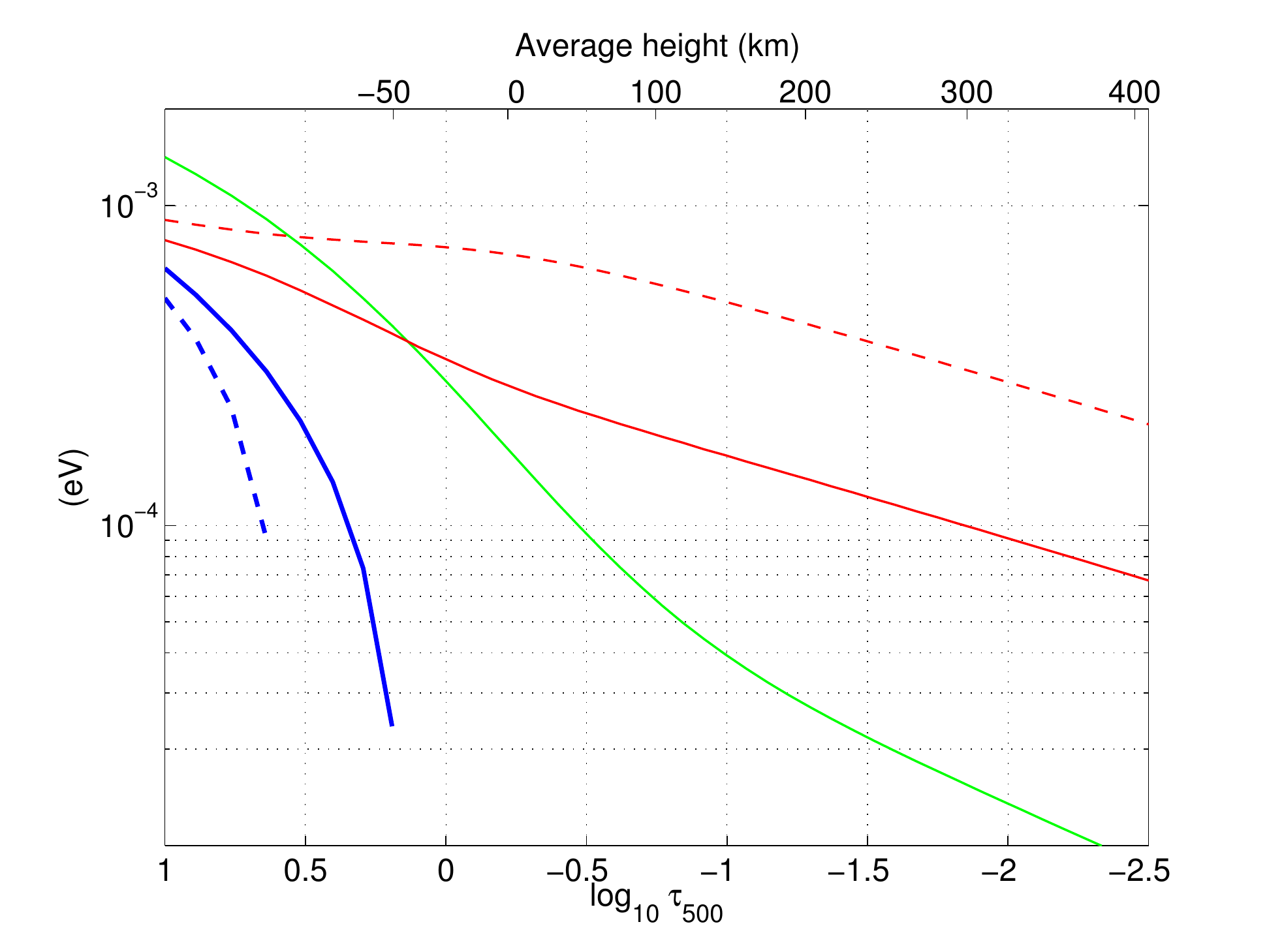}
  \caption{Green: average plasma frequency $\wpl$ (eV) as a function of optical depth at 500 nm (bottom axis) and averaged height (upper axis) in a convection simulation of the Solar atmosphere. Red: contribution of the neutral hydrogen $(\omega \Delta_{||}^{\rm{gas}})^{1/2}$ at photon energies corresponding to $\lambda = 300$\,nm (dashed) and 900 nm (solid). When the difference between the plasma and neutral hydrogen quantities (thick blue lines) is positive, $\wpl^2 - \omega \Delta_{||}^{\mathrm{gas}} > 0$, resonant enhancement may occur. The value of this difference at a given height corresponds to the ALP mass $m_a$ for which resonant conversion takes place. }
\label{reslocation}
\end{figure}

Eq.\ \ref{P0} describes a Lorentzian (Breit-Wigner) distribution in $\Delta_a\propto 1/\omega$ modulated by the $\sin^2$-term, with resonance at the plasma frequency and width $\Delta_{a\gamma}\propto B$. When $\Delta_{||} - \Delta_a$ goes to zero, photon conversion to ALPs is enhanced. We later show that for $m_a \sim 10^{-6}$--$10^{-3}$\,eV, this has the potential to occur in the region of the solar atmosphere where line-formation occurs. The plasma frequency $\wpl$ increases quickly with increasing optical depth. Given the $\wpl^{-4}$ suppression of Eq.~\ref{P0}, this means that conversion is highly suppressed at depths below the resonance depth \hr. There are thus three qualitatively different regions in the solar atmosphere: 1) Below \hr, conversion is suppressed by $\wpl^{-4}$, ensuring that very little conversion occurs at depth; 2) at \hr, resonance occurs: conversion is enhanced and $P_0 \rightarrow 1$ for a small region; and 3) above \hr, $P_0$ becomes approximately constant, providing moderate conversion rates. The existence of region 2) is contingent on the plasma effects being larger than the contribution from neutral gas; otherwise, $\Delta_{||}$ and $- \Delta_a$ are always positive, prohibiting cancellation.

We point out that many references (e.g.\ \cite{Raffelt:1988,Mirizzi:2006zy}) do not include $\D_{||}^{\rm{gas}}$. This is because in situations such as hot stellar interiors or the interstellar medium, the ionisation fraction is so large that $\D_{||}^{\rm{gas}}$ is always subdominant to $\Dpl$. Indeed, the intermediate temperatures encountered in the solar atmosphere make it one of the few places where both the gas refractive index and plasma frequency contribute to similar degrees.  Fig.\ \ref{reslocation} illustrates these effects, showing that resonant conversion may only occur below $\log \tau \sim 0$ (i.e. at larger optical depths). The blue lines show the ALP mass at which resonant conversion may occur, for two different photon wavelengths.

 We now compute the effect of photon-ALP conversion on the photon intensity, for two complementary magnetic field configurations, starting with the case where the size of the magnetic domains $s$ is larger than the scale height over which line formation occurs. After propagation over a distance $\delta z$, photon loss to photon-axion conversion gives the change in intensity \cite{Mirizzi:2006zy}:
\begin{equation}
 I_\gamma(z) = \left(1-\frac{1}{2}P_0(\delta z)\right)I_\gamma(z_0).
\label{largedomlosseq}
\end{equation} 
This assumes that the change of the photon intensity due to radiative transfer is small over the distance $\delta z$ and does not feed back on the calculation of the conversion rate. The linear attenuation coefficient due to ALP conversion is then
\begin{equation}
 -\frac{I_\gamma'(z)}{I_\gamma(z)} \equiv \alpha_\lambda^a = \frac{\Dosc \sin(\Dosc \delta z)}{\frac{\Dosc^2}{\Dag^2}-1 + \cos(\Dosc \delta z)},
 \label{largedomlossrate}
\end{equation} 
where the prime symbol denotes differentiation with respect to $z$. When computing the full radiation transfer, (\ref{largedomlossrate}) must then be added to the linear attenuation due to opacity.

For a line with some mean optical depth of formation $\tau$, the range of heights over which $\alpha_\lambda^a$ will contribute to the line profile is approximately given by the scale height $\Hk$ with which the continuum opacity changes by a factor of $e$, 
\begin{equation}
 \Hk \equiv \left(\frac{1}{\kappa^{\rm c}_\lambda}\frac{\ud}{\ud z}\kappa^{\rm c}_\lambda \right)^{-1}.
\label{hdef}
\end{equation}
This simply states the fact that to a rough approximation, the width of a line's contribution function is given by the opacity scale height.\footnote{More correctly, the scale height of the line-to-continuous opacity \emph{ratio} $\eta_\lambda$.  As this scale height varies on a line-by-line basis, it is not especially useful for estimating effects on abundances as a function of formation height, in the transition-independent manner we do here.} The relevant contribution of $\alpha_\lambda^a$ to the continuum opacity for any line formed at depth $\tau$ can then be estimated by averaging $\alpha_\lambda^a$ over $\Hk$
\begin{equation}
 \langle \alpha_\lambda^a \rangle =  \frac{1}{\Hk}\int_1^{x(\Hk)}\frac{ \ud x}{\left(\frac{\Dosc}{\Dag}\right)^2-1+x},
\label{avgIntegral}
\end{equation} 
where $x(\delta z) = \cos(\Dosc \delta z)$.
Integrating gives
\begin{equation}
\langle \alpha_\lambda^a \rangle =  -\frac{1}{\Hk} \ln\left\{1-\left(\frac{\Dag}{\Dosc}\right)^2\left[1-\cos(\Dosc \Hk)\right] \right\}.
\label{alphaEq}
\end{equation} 
Away from resonance, $\Dosc \gg \Dag$ and the log may be Taylor-expanded about 1, giving
\begin{equation}
\langle \alpha_\lambda^a \rangle =  \frac{1}{\Hk}\left(\frac{\Dag}{\Dosc}\right)^2\left[1-\cos(\Dosc \Hk)\right].
\label{SCLDalpha}
\end{equation} 
Because we are interested in the resonance region as well, we will only make use of the full expression (Eq.~\ref{alphaEq}).

If photons instead propagate through $n$ domains of length $s$ with equal magnetic field strength but random orientation (such that $\delta z = ns$), their intensity is reduced to \cite{Mirizzi:2006zy}:
\begin{equation}
\left[\begin{array}{c}
       I_\gamma(z) \\
       I_a(z)
      \end{array}
\right] =
\left[\begin{array}{c c}
       1-\frac{1}{2}P_0 & P_0 \\
	 \frac{1}{2}P_0 & 1-P_0
      \end{array}
\right]^n 
\left[\begin{array}{c}
       I_\gamma(z_0) \\
       I_a(z_0)
      \end{array}
\right].
\label{nstep}
\end{equation} 
Here $I_a$ refers to the intensity of the ALP field, and $P_0 = P_0(s)$. If the ALP population is initially zero -- we return to the validity of this assumption below -- the photon intensity becomes :
\begin{equation}
 I_\gamma(z) = \frac{1}{3}\left[2+ \left(1-\frac{3}{2}P_0\right)^n\right]I_\gamma(z_0).
\label{Iofz}
\end{equation} 
 If the opacity scale height $\Hk$ (the propagation length of interest) is much larger than the magnetic domain size $s$, we can use the large $n$ limit of Eq.~\ref{Iofz} \cite{Mirizzi:2006zy}
\begin{align}
 P_{\gta}(\delta z) &= \frac{1}{3}\left[1-\exp{\left(-\frac{3P_0(s)\delta z}{2s}\right)}\right] \label{expP0}, \\
 I_\gamma(z) &= \left[1-P_{\gta}(\delta z)\right]I_\gamma(z_0).
\end{align} 
This is valid for a photon passing through a large number $\delta z/s$ of magnetic domains of random orientation, but with identical field amplitudes, $B$. This serves the same function as the averaging procedure (Eq.~\ref{avgIntegral}). Note that $P_0$ is again evaluated at $s$, the size of the magnetic domains. The linear attenuation coefficient is then
\begin{equation}
 \alpha^a_\lambda =-\frac{I'_\gamma(z)}{I_\gamma(z)} = \frac{P_0}{2s}\frac{3}{1+2\,{\rm exp}\left(\frac{3P_0\delta z}{2s}\right)}.
\label{SDfulleq}
\end{equation}
Away from the resonance region, terms that are higher-order in $P_0$ may once more be dropped, yielding:
\begin{align}
						\alpha^a_\lambda & =\frac{P_0}{2s} + O(P_0^2) \\		
						  &\simeq \frac{1}{s}\left(\frac{\Dag}{\Dosc}\right)^2 \left[1 - \cos(\Dosc s)\right].
\label{SCSD}
 \end{align} 
We see that this is just Eq.~\ref{SCLDalpha}, but with the scale height $\Hk$ replaced by the magnetic domain size $s$. For our calculations in Section \ref{sec:abundances}, we use the the full expressions (Eqs.~\ref{alphaEq} and \ref{SDfulleq}) rather than the small $P_0$ approximations.

A final comment should be made on the topic of ALP-photon reconversion. Because this effect is implicit in the definition of $P_0$ (Eq.~\ref{P0}), it is accounted for in the photon-loss expressions Eqs.~\ref{alphaEq} and \ref{SDfulleq}. However, we have assumed that at every height the initial ALP population is zero, \textit{i.e.} $I_a(z_0) = 0$. Although it is not true in general that no axions will reach a given layer from any of the layers below, this is in fact a good approximation for our purposes. At height $z_0$, the ``initial'' ALP population is given to a good approximation by the conditions in the layer below, so $I_a(z_0) = \frac{1}{2}P_0(\delta z)I_\gamma(z_0-\delta z)$.  Therefore, Eq.~\ref{nstep} dictates that the reconversion rate of the initial axions into ``new'' photons is
\begin{equation}
I_{\gamma,\mathrm{rec}}(z) = \frac{1}{2}P_0(\delta z)^2 I_{\gamma}(z_0-\delta z),
\end{equation}
where $I_{\gamma,\mathrm{rec}}$ is the intensity of reconverted photons only. As $P_0 \ll 1$ everywhere but the resonance region, this quantity should  be smaller than the photon-loss rate as long as the scale height $\Hk$ (and therefore $\delta z$) is larger than the width $\delta h_{\rm res}$ of the resonance region. In all cases of interest to us, $\Hk \simeq 100$\,km, whereas $\delta h_{\rm res} \lesssim 15$\,km. 

\begin{figure*}[tbp]
$\begin{array}{cc}
\includegraphics[width=0.5\textwidth]{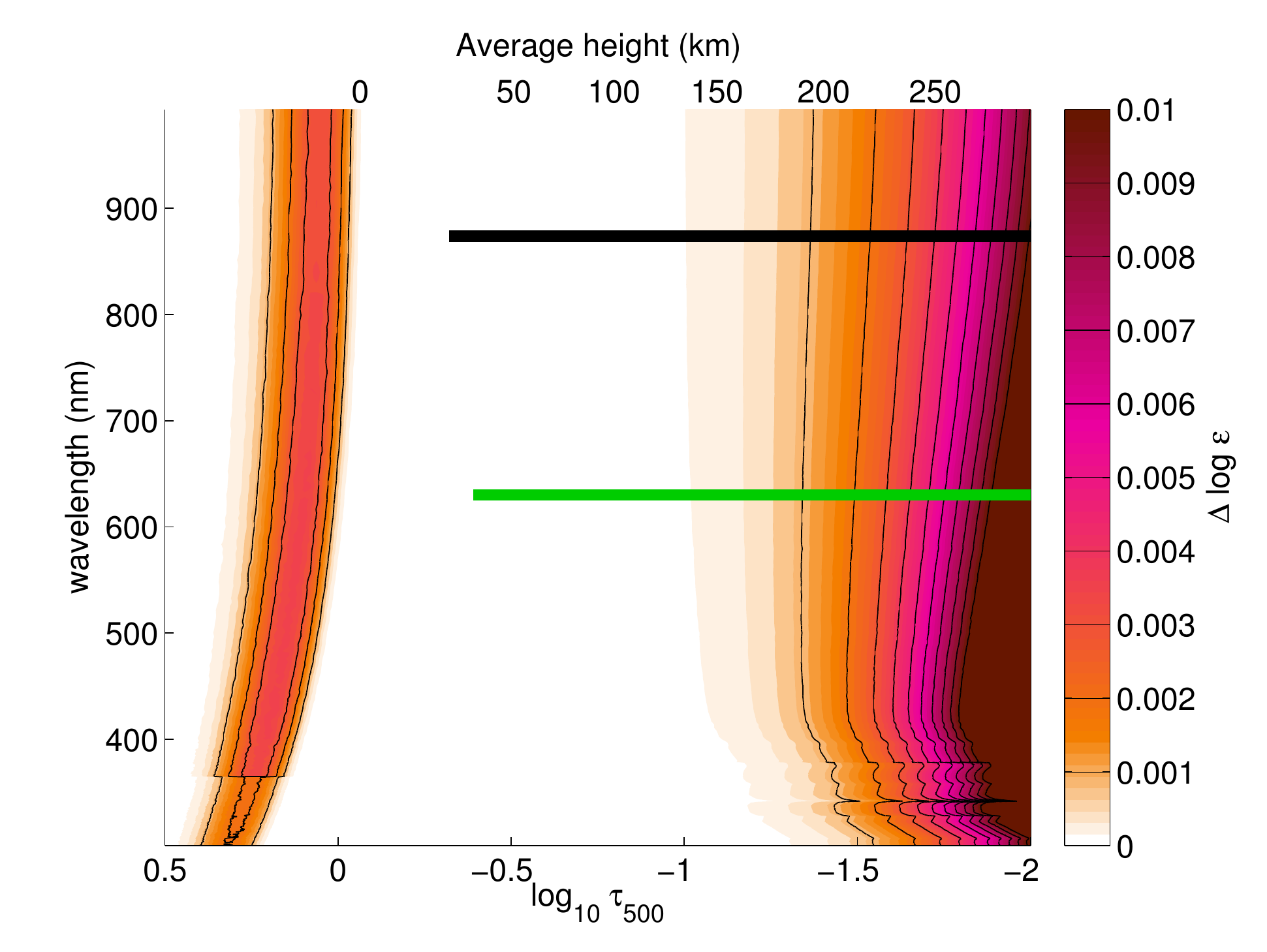} &
\includegraphics[width=0.5\textwidth]{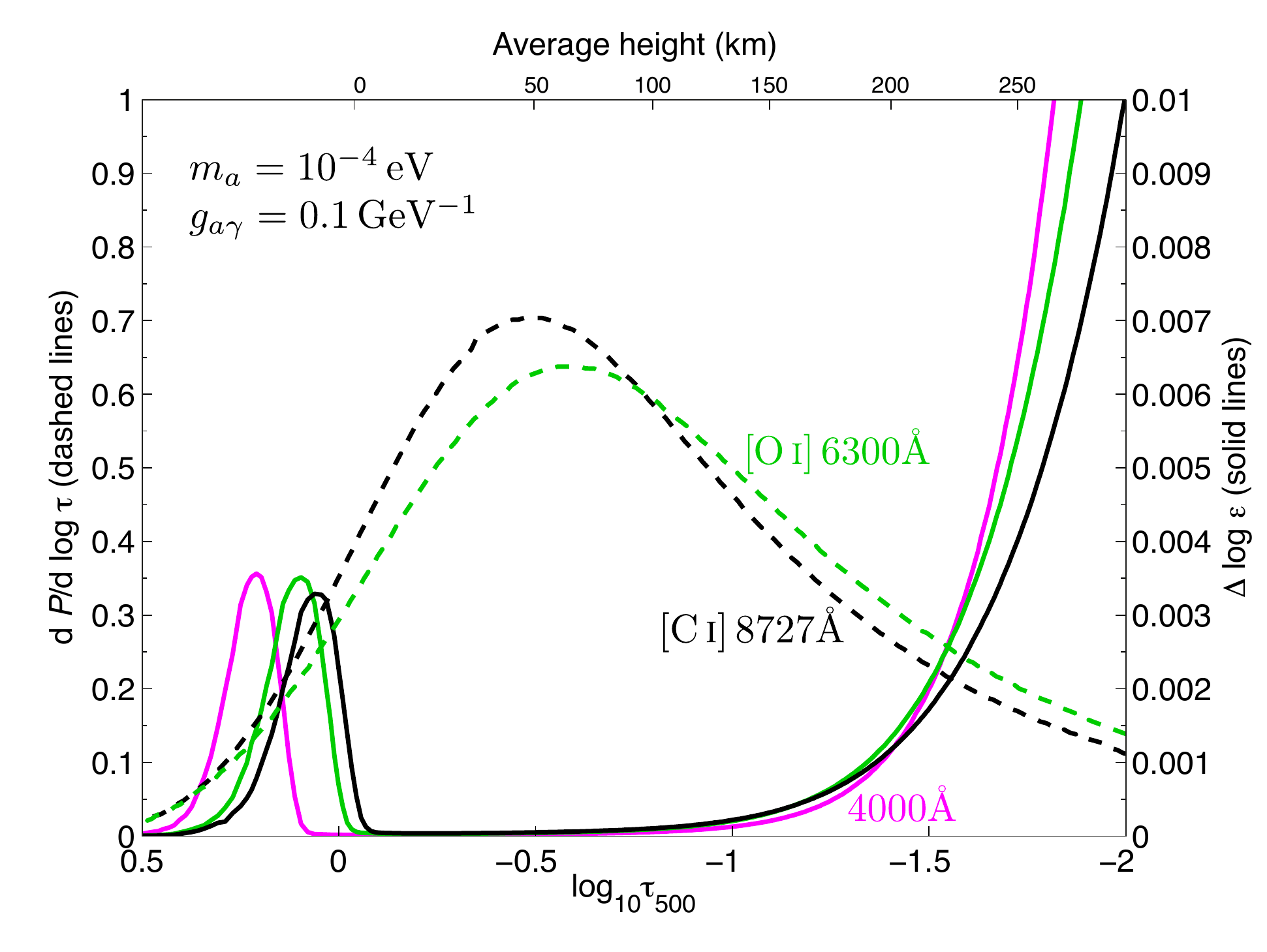}
\end{array}$
\caption{The effect of resonant photon-axion conversion on the effective line widths. Left: Abundance correction effect in the solar atmosphere due to particles with an axion-like coupling, as a function of optical depth $\tau$ and wavelength. Here  $m_a = 10^{-4}$\,eV and we have set $\gag = 0.1$\,\igev, to illustrate the effect, although the allowed value of $\gag$ by observational searches is much lower. Horizontal bars indicate the central 68\% intervals of the line formation contribution functions for the 630.0\,nm [O\,\textsc{i}] (green) and 872.7\,nm [C\,\textsc{i}] (black) forbidden neutral atomic lines. Right: horizontal cross-sections through the left-hand panel for (solid lines) $\lambda= 400$ (magenta), 630\,nm (green) and 875\,nm (black). The peaks are due to resonant-photon-ALP conversion near optical depth $\log \langle \tau \rangle = 0$. Green and black dashed lines are the contribution functions $\ud P/\ud\log\tau$ to the 630.0\,nm [O\,\textsc{i}] and 872.7\,nm [C\,\textsc{i}] absorption lines, respectively. In both cases, $\int \ud P = 1$ with $\tau$ being the monochromatic optical depth at the central wavelength of the respective absorption line (rather than $\tau_{500}$).}
\label{results1}
\end{figure*}

\begin{figure*}[tbp]
$\begin{array}{cc}
\includegraphics[width=0.5\textwidth]{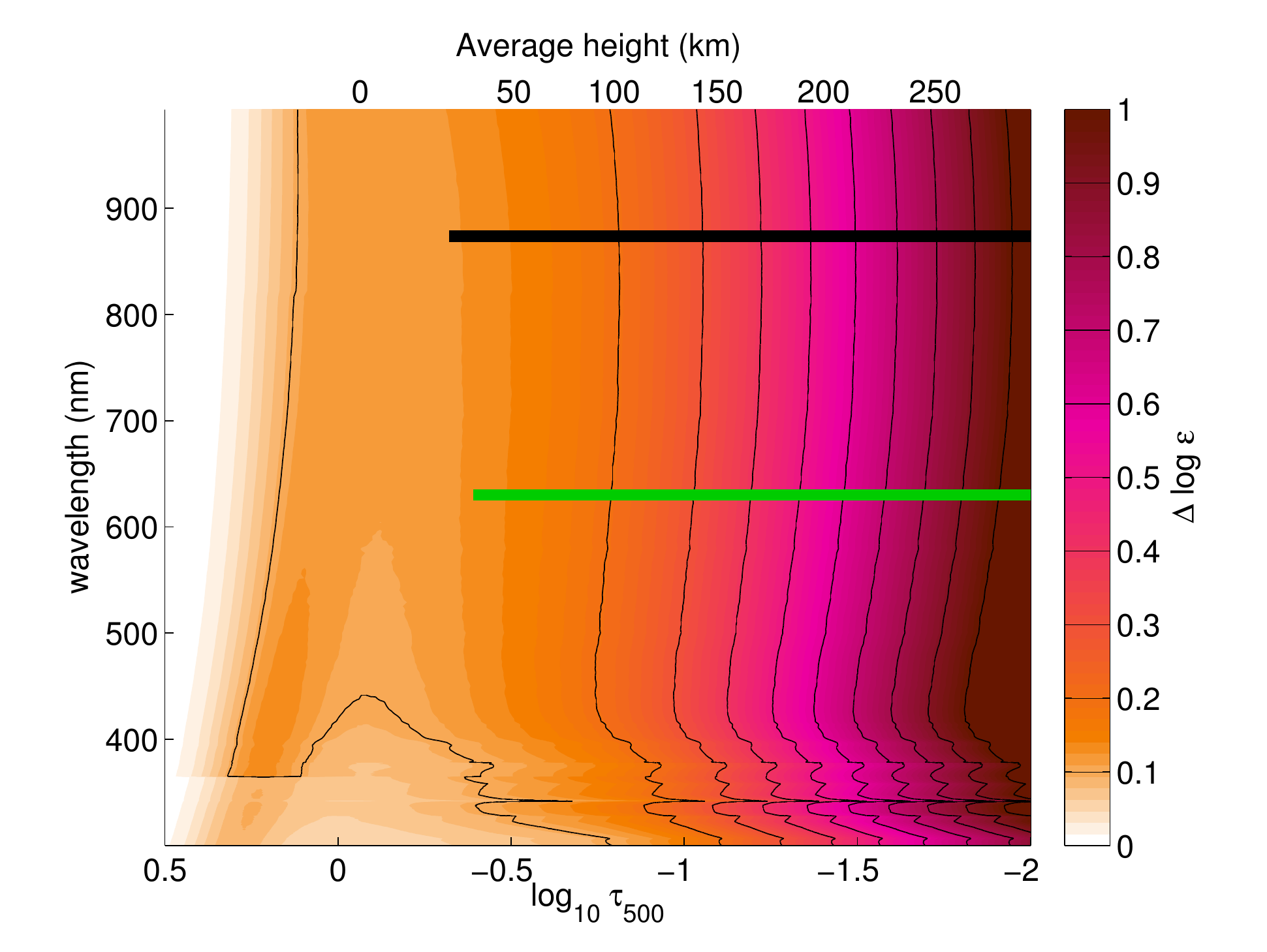} &
\includegraphics[width=0.5\textwidth]{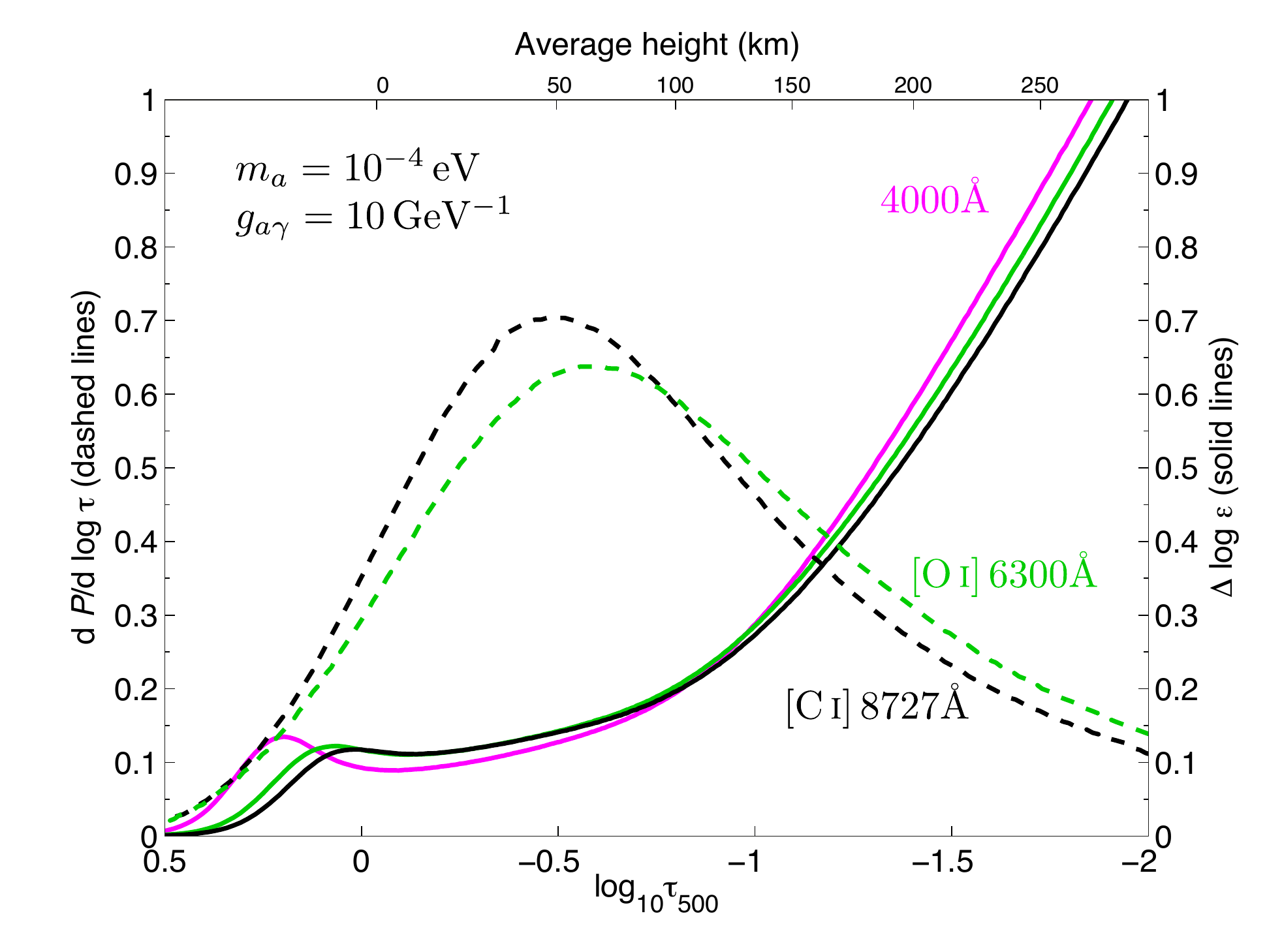}
\end{array}$
\caption{The effect of non-resonant photon-axion conversion, with a coupling large enough to solve the solar abundance problem. Same as Fig.~\ref{results1}, but with $\gag = 10$\,\igev and $m_a = 1\times 10^{-4}$\,eV. Solid lines on right-hand figure are for $\lambda= 400$ (magenta), 630\,nm (green) and 875\,nm (black); dashed lines are the  O\,\textsc{i} and C\,\textsc{i} contribution functions (see Fig. \ref{results1} caption). The conversion probability remains constant throughout the line-forming region of the atmosphere, and the rise in $\Delta \log \epsilon$ is due to the falling bulk opacity with height. We note that the extremely large coupling required is ruled out by several orders of magnitude.}
\label{results2}
\end{figure*}

\section{Effects on solar abundances}
\label{sec:abundances}
To determine the impact of photon-ALP conversion in the photosphere on solar abundances, we calculated the expected correction factor $\Delta\log\epsilon$ (Eq.~\ref{Dloge}) using the 3D hydrodynamic solar model atmosphere from \cite{AGSS}. We computed the ALP-induced opacity $\kappa^a_\lambda$ at wavelengths from 300 to 1000\,nm, at each point in an $(x,y,z)$ grid with resolution $50 \times 50 \times 500$, sampled from the atmosphere simulation. Our extracted grid covers $6\times6$\,Mm, in the horizontal directions, and optical depths in the range $-3\le \log \tau \le 0.5$, in which the majority of line formation occurs and below which conversion is negligible. We did this for 90 snapshots from the simulation, corresponding to 45 minutes in solar time before averaging. All averages presented here are carried out over time and over the undulating
surfaces of particular optical depths to give averages on the $\tau_{500}$-scale.
We averaged the local values of $\kappa^a_\lambda$ over $(x,y,t)$ slices of common optical depth at 500\,nm,
to arrive at expected abundance corrections as a function of wavelength and formation height of arbitrary spectral lines.  Spatial and temporal changes in line profiles due to convective motions occur over the length scales and lifetimes of convective granules, which are approximately 1\,Mm and 10\,min for the Sun \cite{SandN,AspARAA}.  As these scales are significantly smaller than the extent of the simulation, our averaged values should accurately represent the expected effect of ALPs on the mean solar spectrum. 

The continuum opacities include bound-free absorption by the first three
ionisation stages of 15 of the most abundant elements, as computed by the
Opacity Project (OP) and the Iron Project (IP) \cite{OP05}. The absorption by
the anions of H, He, C, N and O is also included, as well as that by the
molecules H$_2$, H$_2^+$, H$_2^-$, OH, CH, H$_2$O and CO$^-$ and the pressure
induced absorption of transient (H\textsc{i}+H\textsc{i})-,
(H\textsc{i}+He\textsc{i})-, (H$_2$+H\textsc{i})-, (H$_2$+He\textsc{i})- and
(H$_2$+H$_2$)-pairs. A complete list with detailed references was presented in
\cite{hayek:Parallel3Dscatter}.  The opacity in the 380--1644\,nm range is
dominated by the bound-free absorption by H$^-$, by free-free absorption of
H\textsc{i} and H$^-$ further into the infra-red and by bound-free absorption
by H\textsc{i} and various metals further into the ultra-violet.

Depending on initial conditions and simulation details, magnetohydrodynamic models of the photosphere predict vertical magnetic field amplitudes in the quiet Sun up to 0.1\,T, with domains ranging from a few hundred to 1000\,km in diameter at line-forming heights \cite{Stein, Cheung:2012cm,Moll:2012mx}. This means that Eq.~\ref{alphaEq} is valid for all our calculations; if $s$ were equal to or smaller than the scale height $\Hk$, Eq.~\ref{SCSD} would be needed instead. The horizontal component of the magnetic field can furthermore be 2 to 5 times larger than the vertical component \cite{Steiner:2009zj}. We therefore take $\langle B \rangle$ = 0.1\,T, but caution that some studies (including \cite{Steiner:2009zj}) favour a more modest 0.001 to 0.01\,T. 

The impact of resonant photon-axion conversion on the effective line opacity is illustrated in Fig.~\ref{results1} with, $m_a = 4\times 10^{-5}$\,eV and a very large value of $\gag = 0.1$\,\igev, giving resonant conversion at a mean optical depth of $\log \langle \tau \rangle \sim 0.2$. We also show the contribution functions for the 630.0\,nm [O\,\textsc{i}] and 872.7\,nm [C\,\textsc{i}] forbidden lines (reproduced from 3D line formation calculations presented in \cite{Caffau08} and \cite{Caffau10}, respectively), indicating the contributions from different atmospheric heights to the formation of these lines.  These two lines are important contributors to the determination of oxygen and carbon abundances \cite{APForbidO,CtoO,AspIV,AspVI,AGSS}. Due to the $\D_{||}^{\rm{gas}} $ contribution, which may be seen as a suppression of the effective photon mass by the index of refraction of neutral hydrogen, resonant conversion within most of the line-forming region is disallowed. We note the sharp rise at low optical depths is because of the sharp \textit{drop} in the opacity at these heights: Fig.~\ref{results1} b) shows the relative change in line strength, reflecting the decrease in continuum opacity; the absolute change at optical depths less than $\sim -2$ has no observable effect. 

If the coupling $\gag$ is pushed to even larger values $\gag > 10$\,\igev, non-resonant conversion can provide corrections of $\Delta \log \epsilon > 0.2$ dex in the centre of the region of line-formation. We illustrate this in Figure \ref{results2}.  Taken naively, the correction factors we present here could offer at least a partial solution to the solar abundance problem, as abundances from all lines would increase relative to their expected values.  This example combination of $m_a$ and $\gag$ would induce the necessary 0.1--0.2\,dex change in abundances from some common C and O indicators; similar effects would be expected with other lines used for abundance determination, as they typically form at similar optical depths.  The exact amount by which the the abundance inferred from each line and species would rise would depend on line choice, line-to-line differences in wavelengths and formation depths, the magnetic field structure and so on.

However, to this point we have not considered the viability of the couplings required, in in view of independent constraints from solar core or laboratory searches for axions and ALPs.  In the following Section we discuss those bounds and the possibility of achieving this effect with real particle models, showing that any observable effect is in fact already ruled out by existing bounds on ALPs.

\section{Discussion}
\label{sec:discussion}

\subsection{ALPs}

There are three types of photon-ALP conversions that can be probed in order to constrain $\gag$ and $m_a$\footnote{The notation $M_{a\gamma} \equiv 1/\gag$ is often used in such discussions.}: 
\begin{enumerate}
\item Conversion in hot dense cores of stars should produce X-ray energy ALPs. Observations of horizontal branch stars constrain the rate of energy loss during their He-burning phase \cite{Raffelt:1999tx}. The Sun itself would lose significant amounts of energy from its core if standard ALPs with couplings large enough to solve the solar abundance problem truly existed, giving rise to observable effects in the frequencies of p-mode (sound wave) oscillations \cite{WeissRaffelt} and the solar neutrino flux \cite{GondoloRaffelt}.  Conversion in the Sun can also be directly probed: pointing an ``axion helioscope'' at the Sun give strict limits on the reconversion of ALPs into X-rays \cite{Arik:2008mq}. The CERN Axion Solar Telescope (CAST) bound on Solar core ALP production and reconversion is $\gag < 9 \times 10^{-11}$\,\igev\ for masses less than $10^{-2}$\,eV.
\item Light shining through wall (LSW) experiments probe reconversion of ALPs produced from laser beams in an intense magnetic field. Current bounds are from the ALPS \cite{Ehret2010149} and GammeV \cite{Chou:2007zzc} experiments. Both shine a high-intensity 532\,nm laser beam at a stopper in a transverse magnetic field, and set bounds based on the (non) observation of reconverted photons behind the stopper.  For masses below 1\,MeV, the ALPS bound is $\gag \lesssim 6 \times 10^{-8}$\,\igev\  whereas the GammeV bound is $\gag \lesssim 3 \times 10^{-7}$\,\igev. 
\item Vacuum polarisation changes from ALP conversion of a linearly polarised infra-red laser beam in a 2.3\,T magnetic field has been probed by the PVLAS experiment \cite{PVLAS}. Vacuum \textit{dichroism} occurs when some of the photons oscillate into ALPs, causing a loss of part of the beam with polarisation parallel to the magnetic field, and thus cause a small beam rotation. Conversely, \textit{birefringence} occurs when ALP oscillation causes a phase lag of one of the polarisation components, causing an ellipticity to develop in the beam polarisation. The strongest bound to date from this effect has been provided by the dichroism limits of PVLAS: $\gag \lesssim 4 \times 10^{-7}$\,\igev \,for ALP masses below $\sim$ 1 meV.
\end{enumerate}

These limits are much too strong to allow standard axions or ALPs to produce an observable effect in the solar atmosphere. Nonetheless, there are classes of models in which some of these bounds do not apply. It was pointed out by \cite{Masso:2005ym} that if the ALP is a composite object analogous to the neutral pion, then the photon-ALP conversion can be suppressed at high energies when one of the photons is virtual. This suppresses ALP production in the hot interiors of stars, alleviating constraint 1. If the ALPs are furthermore short-lived particles, for instance if they decay to a lighter particle that does not interact strongly with the standard model, then constraint 2 may also be overcome. In the latter case, the reconversion discussion in section \ref{sec:abundances} becomes irrelevant. 

However, dichroism constraints from PVLAS cannot be brushed aside even for non-standard models of ALPs that can avoid constraints 1 and 2. Indeed, the photon-loss that would be observed in a dichroism experiment is the same as the effect that would be responsible for photon loss in the line-forming region of the solar atmosphere. If ALPs must respect the PVLAS upper limit on their coupling ($\gag \leq 10^{-7}$\,\igev), then the coupling required to produce an observable effect is suppressed by at least eight orders of magnitudes. We therefore conclude that current limits rule out axions and ALPs as progenitors of any observable effects on abundances derived from the solar photosphere.

\subsection{Chameleons and hidden photons}
In addition to ALPs, there exist further models of particles which mix with the photon in an analogous manner to (\ref{axionFF}), and one could hope that such models may allow constraints to be evaded. We look at two specific models: chameleons, and hidden photons.

The chameleon, first proposed as a model of dark energy \cite{Khoury:2003aq,Khoury:2003rn}, consists of a scalar field $\phi$ with an effective coupling to matter $V(\phi) \sim e^{\beta \phi /M_{\mathrm{Pl}}} \rho$, where $\beta$ is a parameter of the theory, $M_{\mathrm{Pl}}$ is the Planck mass and $\rho$ is the matter density. This gives rise to a particle \cite{Davis09,Brax10,Brax12} whose mass depends on the local density via $m_\phi^2 \propto \rho$. Expanding the exponential coupling, the chameleon's first-order coupling to the photon is:
\begin{equation}
\mathcal{L}_{\phi \gamma} = \frac{\phi}{M_{\phi\gamma}} F_{\mu \nu}F^{\mu \nu},
\end{equation}
where $M_{\phi\gamma} = M_{\mathrm{Pl}}/\beta$ is the mass scale of new physics giving rise to the chameleon field.  The only relevant phenomenological difference with ALPs will be that chameleons couple to photon polarisation which is orthogonal, rather than parallel to, the bulk magnetic field. As we assume random polarisations here, this does not affect conversion in the solar atmosphere. The PVLAS experiment gives similar constraints on the chameleon mixing parameter as it does to the ALP case: $1/M_{\phi \gamma} < 10^{-6}$\,\igev \cite{Brax:2007ak}. 

The hidden photon (or paraphoton) paradigm \cite{Redondo08, AHDM, Bjorken09, Essig11, APEX11} has its origins in an additional $U(1)$ gauge field, with field strength $F'$ which mixes very weakly with the standard model photon via a term
\begin{equation}
\mathcal{L}_{\gamma'\gamma} = \frac{1}{2}\chi F_{\mu \nu}' F^{\mu \nu}. 
\end{equation}
A combination of constraints from the cosmic microwave background, Coulomb-interaction experiments and helioscopic observations from CAST constrain the mixing parameter $\chi \lesssim 10^{-7}$ \cite{Mirizzi:2009iz}. 

Crucially, the form of the coupling to photons means that resonant conversion of photons to either chameleons or hidden photons occurs in the same qualitative way as conversion to ALPs:
\begin{equation}
P_{\gamma \rightarrow \phi_i} \propto \frac{1}{\left(m_{\phi_i}^2 - m_{\gamma,\mathrm{eff}} ^2\right)^2 + \mathrm{const.}}.
\end{equation}
Resonant oscillation is then forbidden for either of these scenarios at optical depths $\log \tau \lesssim 0$, due to the dominance of $\Delta_{||}^{\mathrm{gas}}$. Below such atmospheric heights, we find that the allowed parameter space of either theory does not permit oscillation. The hidden photon mixing parameter is roughly equivalent to $\chi \longleftrightarrow \gag B \omega/m_a^2$, whereas the chameleon coupling can be read directly as $1/M_{\phi\gamma} \longleftrightarrow \gag$. Required parameters in both cases fall far above the current exclusion bounds, guaranteeing that no observable conversion may take place in the solar atmosphere.\footnote{This is true for the most optimistic case in which the density-dependent mass of the chameleon field allows resonance around $\log \tau = 0$. If this is not the case, resonance will of course not occur at all.} 

\section{Conclusions}
\label{sec:conclusion}
We have shown that for appropriate masses and couplings, particles with an axion-like interaction (Eq.~\ref{axionFF}) could in principle resolve the solar abundance problem, by inducing a slight increase in the effective continuum opacity at line-forming heights in the solar atmosphere. This in turn would reduce the computed equivalent widths of solar absorption lines for any given elemental abundance, and could bring inferred photospheric abundances into agreement with helioseismological results. 

However, we found that the coupling necessary to obtain such a drastic change is ruled out by current experimental null results. We have shown that chameleons and hidden photons, which exhibit a similar phenomenological behaviour to the ALP, are also disfavoured as a photospheric solution to the solar abundance problem.  We therefore conclude that new light bosons are not able to provide a photospheric solution to the solar abundance problem.

\section*{Acknowledgements}
It is a pleasure to thank Martin Asplund, Remo Collett, Keshav Dasgupta, Anne Davis, Guy Moore, Javier Redondo and the anonymous referee for helpful comments and conversations. A.C.V. was supported by NSERC, FQRNT and European contracts FP7-PEOPLE-2011-ITN, PITN-GA-2011-289442-INVISIBLES.  P.S. was supported by the Lorne Trottier Chair in Astrophysics, an Institute for Particle Physics Theory Fellowship and a Canadian Government Tri-Agency Banting Fellowship, administered by NSERC. R.T was supported by NASA grants NNX08AI57G and NNX11AJ36G.

\bibliographystyle{JHEP_pat}
\bibliography{solarAxions,CandO,AbuGen,CObiblio,DMbiblio}

\providecommand{\href}[2]{#2}\begingroup\raggedright\begin{thebibliography}{10}

\bibitem{APForbidO}
C.~{Allende Prieto}, D.~L. {Lambert}, and M.~{Asplund}, {\it {The Forbidden
  Abundance of Oxygen in the Sun}},  {\em \apjl} {\bf 556} (2001) L63--L66.

\bibitem{CtoO}
C.~{Allende Prieto}, D.~L. {Lambert}, and M.~{Asplund}, {\it {A Reappraisal of
  the Solar Photospheric C/O Ratio}},  {\em \apjl} {\bf 573} (2002) L137--L140.

\bibitem{AspIV}
M.~{Asplund}, N.~{Grevesse}, A.~J. {Sauval}, C.~{Allende Prieto}, and
  D.~{Kiselman}, {\it {Line formation in solar granulation. IV. [O I], O I and
  OH lines and the photospheric O abundance}},  {\em \aap} {\bf 417} (2004)
  751--768.

\bibitem{AspVI}
M.~{Asplund}, N.~{Grevesse}, A.~J. {Sauval}, C.~{Allende Prieto}, and
  R.~{Blomme}, {\it {Line formation in solar granulation. VI. [C I], C I, CH
  and C$_{2}$ lines and the photospheric C abundance}},  {\em \aap} {\bf 431}
  (2005) 693--705.

\bibitem{AGS05}
M.~{Asplund}, N.~{Grevesse}, and A.~J. {Sauval} in {\em ASP Conf. Ser. 336}
  (T.~G. {Barnes III} and F.~N. {Bash}, eds.), Astron. Soc. Pac., San Francisco
  (2005) 25.

\bibitem{ScottVII}
P.~{Scott}, M.~{Asplund}, N.~{Grevesse}, and A.~J. {Sauval}, {\it {Line
  formation in solar granulation. VII. CO lines and the solar C and O isotopic
  abundances}},  {\em \aap} {\bf 456} (2006) 675--688,
  [\href{http://xxx.lanl.gov/abs/astro-ph/0605116}{{\tt astro-ph/0605116}}].

\bibitem{Melendez08}
J.~{Mel{\'e}ndez} and M.~{Asplund}, {\it {Another forbidden solar oxygen
  abundance: the [O I] 5577 {\AA} line}},  {\em \aap} {\bf 490} (2008)
  817--821, [\href{http://xxx.lanl.gov/abs/0808.2796}{{\tt arXiv:0808.2796}}].

\bibitem{Scott09Ni}
P.~{Scott}, M.~{Asplund}, N.~{Grevesse}, and A.~J. {Sauval}, {\it {On the Solar
  Nickel and Oxygen Abundances}},  {\em \apjl} {\bf 691} (2009) L119--L122,
  [\href{http://xxx.lanl.gov/abs/0811.0815}{{\tt arXiv:0811.0815}}].

\bibitem{AGSS}
M.~{Asplund}, N.~{Grevesse}, A.~J. {Sauval}, and P.~{Scott}, {\it {The chemical
  composition of the Sun}},  {\em \araa} {\bf 47} (2009) 481--522,
  [\href{http://xxx.lanl.gov/abs/0909.0948}{{\tt arXiv:0909.0948}}].

\bibitem{GS98}
N.~{Grevesse} and A.~J. {Sauval}, {\it {Standard Solar Composition}},  {\em
  \ssr} {\bf 85} (1998) 161--174.

\bibitem{Bahcall:2004yr}
J.~N. Bahcall, S.~Basu, M.~Pinsonneault, and A.~M. Serenelli, {\it
  {Helioseismological implications of recent solar abundance determinations}},
  {\em \apj} {\bf 618} (2005) 1049--1056,
  [\href{http://xxx.lanl.gov/abs/astro-ph/0407060}{{\tt astro-ph/0407060}}].

\bibitem{Basu:2004zg}
S.~Basu and H.~Antia, {\it {Constraining solar abundances using
  helioseismology}},  {\em \apj} {\bf 606} (2004) L85,
  [\href{http://xxx.lanl.gov/abs/astro-ph/0403485}{{\tt astro-ph/0403485}}].

\bibitem{Bahcall06}
J.~N. {Bahcall}, A.~M. {Serenelli}, and S.~{Basu}, {\it {10,000 Standard Solar
  Models: A Monte Carlo Simulation}},  {\em \apjs} {\bf 165} (2006) 400--431,
  [\href{http://xxx.lanl.gov/abs/astro-ph/0511337}{{\tt astro-ph/0511337}}].

\bibitem{Yang07}
W.~M. {Yang} and S.~L. {Bi}, {\it {Solar Models with Revised Abundances and
  Opacities}},  {\em \apjl} {\bf 658} (2007) L67--L70,
  [\href{http://xxx.lanl.gov/abs/0805.3644}{{\tt arXiv:0805.3644}}].

\bibitem{Basu08}
S.~{Basu} and H.~M. {Antia}, {\it {Helioseismology and solar abundances}},
  {\em \physrep} {\bf 457} (2008) 217--283,
  [\href{http://xxx.lanl.gov/abs/0711.4590}{{\tt arXiv:0711.4590}}].

\bibitem{Serenelli:2009yc}
A.~Serenelli, S.~Basu, J.~W. Ferguson, and M.~Asplund, {\it {New Solar
  Composition: The Problem With Solar Models Revisited}},  {\em \apj} {\bf 705}
  (2009) L123--L127, [\href{http://xxx.lanl.gov/abs/0909.2668}{{\tt
  arXiv:0909.2668}}].

\bibitem{Asplund08}
M.~{Asplund}, {\it {Does the Sun have a subsolar metallicity?}},  in {\em IAU
  Symposium} (L.~{Deng} and K.~L. {Chan}, eds.) {\bf 252} (2008) 13--26.

\bibitem{Drake05}
J.~J. {Drake} and P.~{Testa}, {\it {The `solar model problem' solved by the
  abundance of neon in nearby stars}},  {\em \nat} {\bf 436} (2005) 525--528,
  [\href{http://xxx.lanl.gov/abs/astro-ph/0506182}{{\tt astro-ph/0506182}}].

\bibitem{Morel08}
T.~{Morel} and K.~{Butler}, {\it {The neon content of nearby B-type stars and
  its implications for the solar model problem}},  {\em \aap} {\bf 487} (2008)
  307--315, [\href{http://xxx.lanl.gov/abs/0806.0491}{{\tt arXiv:0806.0491}}].

\bibitem{Ayres06}
T.~R. {Ayres}, C.~{Plymate}, and C.~U. {Keller}, {\it {Solar Carbon Monoxide,
  Thermal Profiling, and the Abundances of C, O, and Their Isotopes}},  {\em
  \apjs} {\bf 165} (2006) 618--651,
  [\href{http://xxx.lanl.gov/abs/astro-ph/0606153}{{\tt astro-ph/0606153}}].

\bibitem{Ayres08}
T.~R. {Ayres}, {\it {Solar Forbidden Oxygen, Revisited}},  {\em \apj} {\bf 686}
  (2008) 731--740.

\bibitem{Socas07}
H.~{Socas-Navarro} and A.~A. {Norton}, {\it {The Solar Oxygen Crisis: Probably
  Not the Last Word}},  {\em \apjl} {\bf 660} (2007) L153--L156.

\bibitem{Guzik10}
J.~A. {Guzik} and K.~{Mussack}, {\it {Exploring Mass Loss, Low-Z Accretion, and
  Convective Overshoot in Solar Models to Mitigate the Solar Abundance
  Problem}},  {\em \apj} {\bf 713} (2010) 1108--1119,
  [\href{http://xxx.lanl.gov/abs/1001.0648}{{\tt arXiv:1001.0648}}].

\bibitem{Serenelli11}
A.~M. {Serenelli}, W.~C. {Haxton}, and C.~{Pe{\~n}a-Garay}, {\it {Solar Models
  with Accretion. I. Application to the Solar Abundance Problem}},  {\em \apj}
  {\bf 743} (2011) 24, [\href{http://xxx.lanl.gov/abs/1104.1639}{{\tt
  arXiv:1104.1639}}].

\bibitem{Caffau08}
E.~{Caffau}, H.-G. {Ludwig}, {\em et.~al.}, {\it {The photospheric solar oxygen
  project. I. Abundance analysis of atomic lines and influence of atmospheric
  models}},  {\em \aap} {\bf 488} (2008) 1031--1046,
  [\href{http://xxx.lanl.gov/abs/0805.4398}{{\tt arXiv:0805.4398}}].

\bibitem{Centeno08}
R.~{Centeno} and H.~{Socas-Navarro}, {\it {A New Approach to the Solar Oxygen
  Abundance Problem}},  {\em \apjl} {\bf 682} (2008) L61--L64,
  [\href{http://xxx.lanl.gov/abs/0803.0990}{{\tt arXiv:0803.0990}}].

\bibitem{Caffau10}
E.~{Caffau}, H.-G. {Ludwig}, {\em et.~al.}, {\it {The solar photospheric
  abundance of carbon. Analysis of atomic carbon lines with the CO5BOLD solar
  model}},  {\em \aap} {\bf 514} (2010) A92,
  [\href{http://xxx.lanl.gov/abs/1002.2628}{{\tt arXiv:1002.2628}}].

\bibitem{Guzik05}
J.~A. {Guzik}, L.~S. {Watson}, and A.~N. {Cox}, {\it {Can Enhanced Diffusion
  Improve Helioseismic Agreement for Solar Models with Revised Abundances?}},
  {\em \apj} {\bf 627} (2005) 1049--1056,
  [\href{http://xxx.lanl.gov/abs/astro-ph/0502364}{{\tt astro-ph/0502364}}].

\bibitem{Castro07}
M.~{Castro}, S.~{Vauclair}, and O.~{Richard}, {\it {Low abundances of heavy
  elements in the solar outer layers: comparisons of solar models with
  helioseismic inversions}},  {\em \aap} {\bf 463} (2007) 755--758,
  [\href{http://xxx.lanl.gov/abs/astro-ph/0611619}{{\tt astro-ph/0611619}}].

\bibitem{Bahcall05}
J.~N. {Bahcall}, A.~M. {Serenelli}, and S.~{Basu}, {\it {New Solar Opacities,
  Abundances, Helioseismology, and Neutrino Fluxes}},  {\em \apjl} {\bf 621}
  (2005) L85--L88, [\href{http://xxx.lanl.gov/abs/astro-ph/0412440}{{\tt
  astro-ph/0412440}}].

\bibitem{Badnell05}
N.~R. {Badnell}, M.~A. {Bautista}, {\em et.~al.}, {\it {Updated opacities from
  the Opacity Project}},  {\em \mnras} {\bf 360} (2005) 458--464,
  [\href{http://xxx.lanl.gov/abs/astro-ph/0410744}{{\tt astro-ph/0410744}}].

\bibitem{Christensen09}
J.~{Christensen-Dalsgaard}, M.~P. {di Mauro}, G.~{Houdek}, and F.~{Pijpers},
  {\it {On the opacity change required to compensate for the revised solar
  composition}},  {\em \aap} {\bf 494} (2009) 205--208,
  [\href{http://xxx.lanl.gov/abs/0811.1001}{{\tt arXiv:0811.1001}}].

\bibitem{Arnett05}
D.~{Arnett}, C.~{Meakin}, and P.~A. {Young}, {\it {The Lambert Problem}},  in
  {\em Cosmic Abundances as Records of Stellar Evolution and Nucleosynthesis}
  (T.~G. {Barnes}, III and F.~N. {Bash}, eds.) {\bf 336} (2005) 235.

\bibitem{Charbonnel05}
C.~{Charbonnel} and S.~{Talon}, {\it {Influence of Gravity Waves on the
  Internal Rotation and Li Abundance of Solar-Type Stars}},  {\em \science}
  {\bf 309} (2005) 2189--2191,
  [\href{http://xxx.lanl.gov/abs/astro-ph/0511265}{{\tt astro-ph/0511265}}].

\bibitem{Frandsen10}
M.~T. {Frandsen} and S.~{Sarkar}, {\it {Asymmetric Dark Matter and the Sun}},
  {\em \prl} {\bf 105} (2010) 011301,
  [\href{http://xxx.lanl.gov/abs/1003.4505}{{\tt arXiv:1003.4505}}].

\bibitem{Taoso10}
M.~{Taoso}, F.~{Iocco}, G.~{Meynet}, G.~{Bertone}, and P.~{Eggenberger}, {\it
  {Effect of low mass dark matter particles on the Sun}},  {\em \prd} {\bf 82}
  (2010) 083509, [\href{http://xxx.lanl.gov/abs/1005.5711}{{\tt
  arXiv:1005.5711}}].

\bibitem{Cumberbatch10}
D.~T. {Cumberbatch}, J.~A. {Guzik}, J.~{Silk}, L.~S. {Watson}, and S.~M.
  {West}, {\it {Light WIMPs in the Sun: Constraints from helioseismology}},
  {\em \prd} {\bf 82} (2010) 103503,
  [\href{http://xxx.lanl.gov/abs/1005.5102}{{\tt arXiv:1005.5102}}].

\bibitem{Zioutas:2007xk}
K.~Zioutas, Y.~Semertzidis, and T.~Papaevangelou, {\it {Overlooked
  astrophysical signatures of axion(-like) particles}},
  \href{http://xxx.lanl.gov/abs/astro-ph/0701627}{{\tt astro-ph/0701627}}.

\bibitem{PQ}
R.~D. {Peccei} and H.~R. {Quinn}, {\it {CP conservation in the presence of
  pseudoparticles}},  {\em \prl} {\bf 38} (1977) 1440--1443.

\bibitem{Raffelt19901}
G.~G. Raffelt, {\it Astrophysical methods to constrain axions and other novel
  particle phenomena},  {\em Physics Reports} {\bf 198} (1990) 1 -- 113.

\bibitem{vanBibber1987}
K.~Van~Bibber, N.~R. Dagdeviren, S.~E. Koonin, A.~K. Kerman, and H.~N. Nelson,
  {\it Proposed experiment to produce and detect light pseudoscalars},  {\em
  Phys. Rev. Lett.} {\bf 59} (1987) 759--762.

\bibitem{Duffy:2009ig}
L.~D. Duffy and K.~van Bibber, {\it {Axions as Dark Matter Particles}},  {\em
  New J.Phys.} {\bf 11} (2009) 105008,
  [\href{http://xxx.lanl.gov/abs/0904.3346}{{\tt arXiv:0904.3346}}].

\bibitem{Asztalos:2006kz}
S.~J. Asztalos, L.~J. Rosenberg, K.~van Bibber, P.~Sikivie, and K.~Zioutas,
  {\it {Searches for astrophysical and cosmological axions}},  {\em
  Ann.Rev.Nucl.Part.Sci.} {\bf 56} (2006) 293--326.

\bibitem{Svrcek:2006yi}
P.~Svrcek and E.~Witten, {\it {Axions In String Theory}},  {\em JHEP} {\bf
  0606} (2006) 051, [\href{http://xxx.lanl.gov/abs/hep-th/0605206}{{\tt
  hep-th/0605206}}].

\bibitem{Dasgupta:2008hb}
K.~Dasgupta, H.~Firouzjahi, and R.~Gwyn, {\it {On The Warped Heterotic Axion}},
   {\em JHEP} {\bf 0806} (2008) 056,
  [\href{http://xxx.lanl.gov/abs/0803.3828}{{\tt arXiv:0803.3828}}].

\bibitem{Rutten}
R.~J. {Rutten}, {\em {Radiative Transfer in Stellar Atmospheres}}.
\newblock Lecture Notes, Utrecht University, 8th Edition,
  \url{http://www.astro.uu.nl/~rutten}, 2003.

\bibitem{Raffelt:1988}
G.~G. Raffelt, {\it Plasmon decay into low-mass bosons in stars},  {\em \prd}
  {\bf 37} (1988) 1356--1359.

\bibitem{Raffelt:2006cw}
G.~G. Raffelt, {\it {Astrophysical axion bounds}},  {\em \lnp} {\bf 741} (2008)
  51--71, [\href{http://xxx.lanl.gov/abs/hep-ph/0611350}{{\tt
  hep-ph/0611350}}].

\bibitem{Raffelt:1987im}
G.~Raffelt and L.~Stodolsky, {\it {Mixing of the Photon with Low Mass
  Particles}},  {\em \prd} {\bf 37} (1988) 1237.

\bibitem{Mirizzi:2006zy}
A.~Mirizzi, G.~G. Raffelt, and P.~D. Serpico, {\it {Photon-axion conversion in
  intergalactic magnetic fields and cosmological consequences}},  {\em \lnp}
  {\bf 741} (2008) 115--134,
  [\href{http://xxx.lanl.gov/abs/astro-ph/0607415}{{\tt astro-ph/0607415}}].

\bibitem{bartlett:HIpolariz}
M.~L. Bartlett and E.~A. Power, {\it Relativistic corrections to $s_{-2}$ for
  atomic hydrogen},  {\em J. Phys. A} {\bf 2} (1969) 419--426.

\bibitem{SandN}
R.~F. {Stein} and {\AA}.~{Nordlund}, {\it {Simulations of Solar Granulation. I.
  General Properties}},  {\em \apj} {\bf 499} (1998) 914.

\bibitem{AspARAA}
M.~{Asplund}, {\it {New Light on Stellar Abundance Analyses: Departures from
  LTE and Homogeneity}},  {\em \araa} {\bf 43} (2005) 481--530.

\bibitem{OP05}
N.~R. Badnell, M.~A. Bautista, {\em et.~al.}, {\it Up-dated opacities from the
  {Opacity} {Project}},  {\em MNRAS} {\bf 360} (2005) 458--464.

\bibitem{hayek:Parallel3Dscatter}
W.~Hayek, M.~Asplund, {\em et.~al.}, {\it Radiative transfer with scattering
  for domain-decomposed {3D} {MHD} simulations of cool stellar atmospheres.
  {N}umerical methods and application to the quiet, non-magnetic, surface of a
  solar-type star},  {\em A\&A} {\bf 517} (2010) A49.

\bibitem{Stein}
R.~F. Stein and Ã.~Nordlund, {\it Solar small-scale magnetoconvection},  {\em
  The Astrophysical Journal} {\bf 642} (2006) 1246.

\bibitem{Cheung:2012cm}
M.~C. Cheung and R.~H. Cameron, {\it {Magnetohydrodynamics of the Weakly
  Ionized Solar Photosphere}},  {\em Astrophys.J.} {\bf 750} (2012) 6,
  [\href{http://xxx.lanl.gov/abs/1202.1937}{{\tt arXiv:1202.1937}}].

\bibitem{Moll:2012mx}
R.~Moll, R.~Cameron, and M.~Schussler, {\it {Vortices, shocks, and heating in
  the solar photosphere: effect of a magnetic field}},
  \href{http://xxx.lanl.gov/abs/1201.5981}{{\tt arXiv:1201.5981}}.

\bibitem{Steiner:2009zj}
O.~Steiner, R.~Rezaei, R.~Schlichenmaier, W.~Schaffenberger, and
  S.~Wedemeyer-Bohm, {\it {The Horizontal Magnetic Field of the Quiet Sun:
  Numerical Simulations in Comparison to Observations with Hinode}},
  \href{http://xxx.lanl.gov/abs/0904.2030}{{\tt arXiv:0904.2030}}.

\bibitem{Raffelt:1999tx}
G.~G. Raffelt, {\it {Particle physics from stars}},  {\em \arnps} {\bf 49}
  (1999) 163--216, [\href{http://xxx.lanl.gov/abs/hep-ph/9903472}{{\tt
  hep-ph/9903472}}].

\bibitem{WeissRaffelt}
H.~{Schlattl}, A.~{Weiss}, and G.~{Raffelt}, {\it {Helioseismological
  constraint on solar axion emission}},  {\em Astroparticle Physics} {\bf 10}
  (1999) 353--359, [\href{http://xxx.lanl.gov/abs/hep-ph/9807476}{{\tt
  hep-ph/9807476}}].

\bibitem{GondoloRaffelt}
P.~{Gondolo} and G.~G. {Raffelt}, {\it {Solar neutrino limit on axions and
  keV-mass bosons}},  {\em \prd} {\bf 79} (2009) 107301,
  [\href{http://xxx.lanl.gov/abs/0807.2926}{{\tt arXiv:0807.2926}}].

\bibitem{Arik:2008mq}
{\bf CAST} Collaboration, E.~Arik {\em et.~al.}, {\it {Probing eV-scale axions
  with CAST}},  {\em \jcap} {\bf 0902} (2009) 008,
  [\href{http://xxx.lanl.gov/abs/0810.4482}{{\tt arXiv:0810.4482}}].

\bibitem{Ehret2010149}
K.~Ehret, M.~Frede, {\em et.~al.}, {\it New alps results on hidden-sector
  lightweights},  {\em \plb} {\bf 689} (2010) 149 -- 155.

\bibitem{Chou:2007zzc}
{\bf GammeV (T-969)} Collaboration, A.~S.~. Chou {\em et.~al.}, {\it {Search
  for axion-like particles using a variable baseline photon regeneration
  technique}},  {\em \prl} {\bf 100} (2008) 080402,
  [\href{http://xxx.lanl.gov/abs/0710.3783}{{\tt arXiv:0710.3783}}].

\bibitem{PVLAS}
{\bf PVLAS} Collaboration, M.~Bregant, G.~Cantatore, {\em et.~al.}, {\it Limits
  on low energy photon-photon scattering from an experiment on magnetic vacuum
  birefringence},  {\em \prd} {\bf 78} (2008) 032006.

\bibitem{Masso:2005ym}
E.~Masso and J.~Redondo, {\it {Evading astrophysical constraints on axion-like
  particles}},  {\em \jcap} {\bf 0509} (2005) 015,
  [\href{http://xxx.lanl.gov/abs/hep-ph/0504202}{{\tt hep-ph/0504202}}].

\bibitem{Khoury:2003aq}
J.~Khoury and A.~Weltman, {\it {Chameleon fields: Awaiting surprises for tests
  of gravity in space}},  {\em Phys.Rev.Lett.} {\bf 93} (2004) 171104,
  [\href{http://xxx.lanl.gov/abs/astro-ph/0309300}{{\tt astro-ph/0309300}}].

\bibitem{Khoury:2003rn}
J.~Khoury and A.~Weltman, {\it {Chameleon cosmology}},  {\em Phys.Rev.} {\bf
  D69} (2004) 044026, [\href{http://xxx.lanl.gov/abs/astro-ph/0309411}{{\tt
  astro-ph/0309411}}].

\bibitem{Davis09}
A.-C. {Davis}, C.~A.~O. {Schelpe}, and D.~J. {Shaw}, {\it {Effect of a
  chameleon scalar field on the cosmic microwave background}},  {\em \prd} {\bf
  80} (2009) 064016, [\href{http://xxx.lanl.gov/abs/0907.2672}{{\tt
  arXiv:0907.2672}}].

\bibitem{Brax10}
P.~{Brax} and K.~{Zioutas}, {\it {Solar chameleons}},  {\em \prd} {\bf 82}
  (2010) 043007, [\href{http://xxx.lanl.gov/abs/1004.1846}{{\tt
  arXiv:1004.1846}}].

\bibitem{Brax12}
P.~{Brax}, A.~{Lindner}, and K.~{Zioutas}, {\it {Detection prospects for solar
  and terrestrial chameleons}},  {\em \prd} {\bf 85} (2012) 043014,
  [\href{http://xxx.lanl.gov/abs/1110.2583}{{\tt arXiv:1110.2583}}].

\bibitem{Brax:2007ak}
P.~Brax, C.~van~de Bruck, and A.-C. Davis, {\it {Compatibility of the
  chameleon-field model with fifth-force experiments, cosmology, and PVLAS and
  CAST results}},  {\em Phys.Rev.Lett.} {\bf 99} (2007) 121103,
  [\href{http://xxx.lanl.gov/abs/hep-ph/0703243}{{\tt hep-ph/0703243}}].

\bibitem{Redondo08}
J.~{Redondo}, {\it {Helioscope bounds on hidden sector photons}},  {\em \jcap}
  {\bf 7} (2008) 8, [\href{http://xxx.lanl.gov/abs/0801.1527}{{\tt
  arXiv:0801.1527}}].

\bibitem{AHDM}
N.~{Arkani-Hamed}, D.~P. {Finkbeiner}, T.~R. {Slatyer}, and N.~{Weiner}, {\it
  {A theory of dark matter}},  {\em \prd} {\bf 79} (2009) 015014,
  [\href{http://xxx.lanl.gov/abs/0810.0713}{{\tt arXiv:0810.0713}}].

\bibitem{Bjorken09}
J.~D. {Bjorken}, R.~{Essig}, P.~{Schuster}, and N.~{Toro}, {\it {New
  fixed-target experiments to search for dark gauge forces}},  {\em \prd} {\bf
  80} (2009) 075018, [\href{http://xxx.lanl.gov/abs/0906.0580}{{\tt
  arXiv:0906.0580}}].

\bibitem{Essig11}
R.~{Essig}, P.~{Schuster}, N.~{Toro}, and B.~{Wojtsekhowski}, {\it {An electron
  fixed target experiment to search for a new vector boson A' decaying to e
  $^{+}$ e $^{-}$}},  {\em Journal of High Energy Physics} {\bf 2} (2011) 9,
  [\href{http://xxx.lanl.gov/abs/1001.2557}{{\tt arXiv:1001.2557}}].

\bibitem{APEX11}
S.~{Abrahamyan}, Z.~{Ahmed}, {\em et.~al.}, {\it {Search for a New Gauge Boson
  in Electron-Nucleus Fixed-Target Scattering by the APEX Experiment}},  {\em
  Physical Review Letters} {\bf 107} (2011) 191804,
  [\href{http://xxx.lanl.gov/abs/1108.2750}{{\tt arXiv:1108.2750}}].

\bibitem{Mirizzi:2009iz}
A.~Mirizzi, J.~Redondo, and G.~Sigl, {\it {Microwave Background Constraints on
  Mixing of Photons with Hidden Photons}},  {\em JCAP} {\bf 0903} (2009) 026,
  [\href{http://xxx.lanl.gov/abs/0901.0014}{{\tt arXiv:0901.0014}}].

\end{thebibliography}\endgroup

\end{document}